# Hybrid Plant Models Call for a Different Plant Modelling Paradigm and a New Generation of Software

## (Heresy in the land of moles, fractions, & rigorous physical properties)

Vladimir Mahalec

Department of Chemical Engineering, McMaster University, 1280 Main St. West, Hamilton, ON L8S 4LS, Canada

## Abstract

This paper is an invitation to the process systems engineering community to change the paradigm for process plants. The goal is to achieve much easier convergence while retaining accuracy on par with the rigorous models. Accurate plant models of existing plants can be linear or much less nonlinear if they are based on mass component flows and stream properties per unit mass properties instead of molar flows and mole fractions. Accurate stream properties per unit mass can be calculated at stream specific conditions by linear approximations which in many instances eliminates mole fraction-based flash calculations. Hybrid data-driven node models fit naturally in this paradigm, since they used measured data, which is either in mass or in volumetric units, but never in moles. Instantiation of models at all levels of abstraction (planning, scheduling, optimization, and control models) from the same plant topology representation will ensure inheritance of solutions from mass-only to mass-and-energy to mass-and-energy-and-stream-properties, thereby ensuring consistency of solutions between these models. None of the existing software provides inheritance between different levels of plant abstraction (i.e. inheritance between models for different business applications) or different levels of abstractions per plant sections or per time periods, which motivates this exposition.

**Keywords**: hybrid plant modelling paradigm, unified plant models for all decision making, model instantiations from the same plant topology, no mole flows and no mole fractions, component mass flows, update of thermophysical properties at local conditions

---

*Corresponding author: mahalec@mcmaster.ca

## 1. Introduction and review of current modelling paradigms

This paper is a call to change the current plant modelling paradigms to a different, unified paradigm which attains accuracy on par with the rigorous plant modelling while eliminating many plant-level nonlinearity that appear in the current rigorous modelling paradigm. It is proposed that the process flow diagram containing all equipment in the plant is the unified representation of the plant, "mother of all plant models". Plant models of different abstraction levels are then instantiated by employing different levels of stream and node abstractions.

Optimization of plant models at different abstraction levels leads to solutions which are in some proximity of the true optimum; the distance from the true optimum depends on the level of model abstraction. Fig. 1 represents conceptual minimization of operating costs of a hypothetical plant. Optimization based on mass balances only has the lowest cost since it does not account for energy costs. If one adds energy balances at fixed stream unit enthalpies, the minimal cost increases and the optimum may lie at somewhat different point. Inclusion of energy balances brings us to the vicinity of the true optimum, since the bulk material and energy costs determine the vicinity of the true optimum. If we go one step further and include local approximation of stream unit enthalpy (and other properties), the optimum solution is likely to change but not very far. We have a somewhat better (arguably not very different) better solution at the expense of introducing nonlinear energy balance equations. Going even further by including rigorous thermophysical property calculations may give some further accuracy, but the overall gain is likely not to be significant. In other words, moving from approximate calculation of physical properties to rigorous calculation of these properties most of the time represents a minor wiggle relative to the solution.

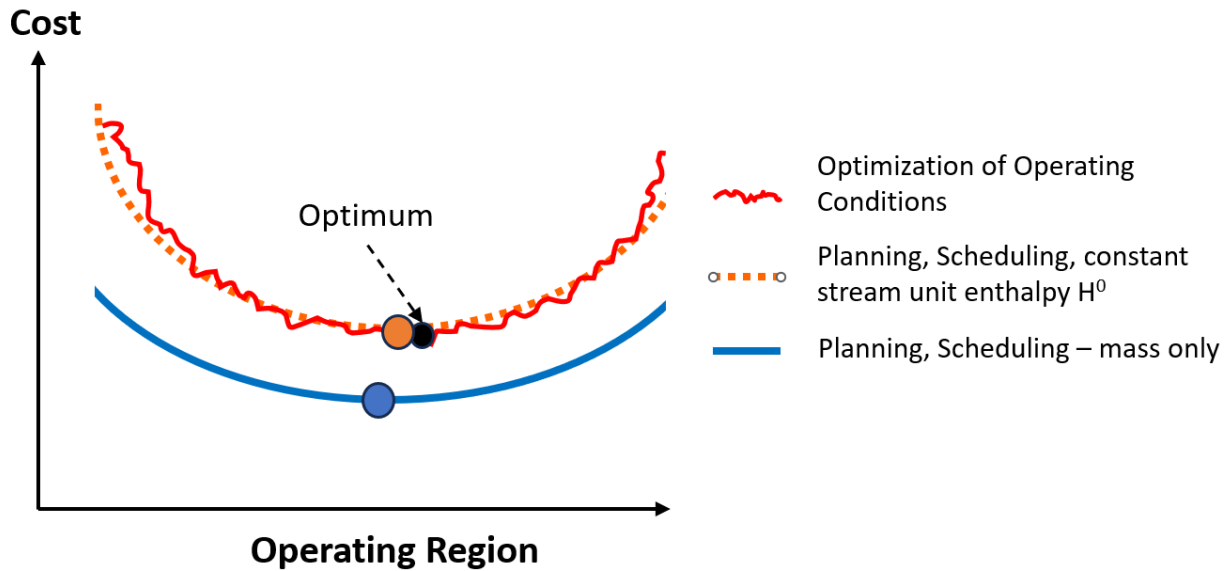

Figure 1. Optimization solutions at different abstraction levels

Please note that moving from coarser (mass balance only) to finer (mass and energy balances) abstractions results in solutions at a coarser level that can be used as initial points for the more detailed abstraction of the model. The unified modeling paradigm proposed by this work enables such inheritance through construction of composite solution algorithms that successively solve the models that are based on the same plant topology.

In addition to plant models being instantiated at different levels of stream abstraction from one business application to another, different levels of stream abstraction are also needed at different sections of a flowsheet in order to make them easier to solve and meet the business purpose for which the model is built. Similarly, node models need to be able to be instantiated at different levels of abstraction (e.g. material balance only, or material end energy balances, or...).

It is proposed that process streams and material inventories be measured in mass units which is the measurement environment in the plants. This enables integration of hybrid equipment models without (to some extent error prone) conversion to moles and the use simplified thermodynamic calculations for plant wide optimization, scheduling, and integration with control. Proposed unifying paradigm eliminates dichotomies between three current plant modelling paradigms.

Since the late 1960s plant modelling for process design and optimization of operating conditions has been based on describing process streams by mole fractions of individual components, total molar flow, temperature and pressure (moles & fractions paradigm). That enabled use of thermodynamic methods to calculate phase conditions of any stream, as well as development of first-principles based models of unit operations. Resulting plant models are highly nonlinear; they require good initialization of individual unit models and of streams in the flowsheet model in order to converge. Sequential modular calculational algorithm, originating in 1960s, is still the prevailing solution procedure due to its ability to converge (slowly) almost any flowsheet. "Equation oriented" simultaneous solution of all flowsheet equations, commercially available since 1990s, has enabled optimization of entire plant models, provided that excellent initialization is provided. Most advanced applications based on this approach have been real-time optimization of large petrochemical plants (e.g. simultaneous RTO of two ethylene plants via model containing more than 500k equations which was implemented by AspenTech in 1990s). Many thousands of person-years have been invested in development of sophisticated process modelling software, enabling relatively straightforward process model configuration and solution (e.g. AspenPlus, Pro/II, Aspen HYSYS). Despite these efforts, models based on this paradigm are constrained to process design and single time-period optimization of operating conditions due to difficulties in converging them as their size grows. It is not possible to use them for multi-period planning or for scheduling.

On the other hand, production planning started to use mathematical models in the early 1950s by using linear programming models to blend gasoline and then proceeded to address the entire refinery production planning (Bodington 1990). These models were based on bulk properties and total stream flows (volumetric or mass). Operation of process units was represented by linearized models, which enabled solution of large-scale multi-period planning models. Planning models today still use the same paradigm, with some modifications to accommodate nonlinear models. Similarly, production scheduling (Weaver 2006) employs models based on bulk flows and properties of streams.

In the late 1970s dynamic matrix controller (DMC) was introduced (Cutler 1979) bringing data-driven plant models into industrial practice. That is the third widely deployed plant modeling paradigm. DMC controller models are the most successful, most widely applied data driven models up to now.

Current lack of unified plant modelling paradigm requires that plant models be built at different levels of abstraction (within different paradigms) and does not permit smooth transition from one level of abstraction to another, which in creates gaps in decision making between planning, scheduling, RTO, and control.

During the last one or two decades more and more attention has been devoted to hybrid models which combine first principles with data-driven models. These efforts have accelerated recently with wide attention given to machine learning and AI methods in the process systems field. Not surprisingly, companies producing plant modelling software view hybrid unit models as just another variation of equipment models and promote that they be integrated as node models in moles & fractions paradigm flowsheet models. Unfortunately, that will not enable moles & fractions software to be used for anything else beyond what it does already.

In the remainder of this paper, we will first discuss various plant level abstractions and their use in decision making processes and then follow it with examples showing that uniform plant level abstraction is not appropriate and there needs to be options to instantiate different parts of a plant at different levels of abstraction. Section 2 discusses changing stream and material representation for total molar flow and mole fractions ("moles & fractions" paradigm) to mass component flow ("mass & flows" paradigm). This enables development of hybrid process unit models that are linear or less nonlinear than mole fraction-based unit models (e.g. reactors). In addition, it or eliminates a significant number of bilinear terms. Section 3 introduces computation of energy balances based on unit enthalpy per mass instead of unit enthalpy per mole since the former is less sensitive to changes in stream composition. Energy balances computed from unit enthalpies per mass at normal operating conditions of streams are introduced, which leads to a linear mass and energy balances (if node models are linear). Such solutions are in the vicinity of the optimum so long as the plant operates in the normal operating region. Using mass instead of moles makes it possible to use local approximations of the bulk physical properties ad still have accuracy close to rigorous property calculations. This avoids stream flash calculations for every stream (as done in rigorous simulation software) and eliminates many nonlinear terms from the plant model. Section 4 discusses consistency between different levels of plant model abstractions and their relationships to the business process, decision making tools (e.g. planning, scheduling, RTO). Section 5 presents composite algorithms for solving more detailed plant models (e.g. mass and energy balances) by starting from simpler, higher abstraction models (e.g. mass balances). Moving from simplest plant models to more detailed models that rely on simplified or rigorous thermophysical properties calculations provides solutions required for planning or scheduling or detailed plant optimization, in addition to providing good starting points for more detailed solutions. Requirement to model different plant sections by different abstraction levels are presented in Section 6  discusses  novel integration of solutions to different decision-making models by employing different incarnations of node models at different time periods in multi-period models. This leads to integrated solutions of e.g. real-time optimization and process control, or integration of planning and scheduling, etc.  The need for a new generation of plant modelling and optimization software is discussed in Section 7.  Conclusions are presented in Section 8.

## 2. Moles & Fractions vs. Component Mass Flows Mass Balance Models

This section presents typical node models corresponding to moles & fractions and to mass & component flows paradigms to show reduction in nonlinear terms that is accomplished by this change.  Variations of mass & component flows models (or volume & component flows) are often used in planning models. (Hutchison 1974) proposed to use component molar flows in plant simulations.

Switching from moles to mass units switch to advocated because it enhances accuracy of simplified thermodynamic calculations as shown in Section 3.

Equations (1) to (17) compare total molar flow & fractions paradigm & to total mass flow and mass flows of components paradigm.

**Stream Mixing**

*Fraction-based mixing model* represented by Eq. (1) to (4)  contains NIN*NC (eq. 1) plus NC bilinear terms (eq. 4).

$$F_i x_{i,j} - F_{i,j} = 0 \qquad i=1,...,NIN; j=1,...,NC \qquad (1)$$

$$\sum_{i=1}^{NIN} F_{i,j} - F_{out,j} = 0 \tag{2}$$

$$\sum_{j=1}^{NC} F_{out,j} - F_{out} = 0 \tag{3}$$

$$F_{out} x_{out,j} - F_{out,j} = 0 \tag{4}$$

*Flow-based mixing model* represented by eqs. (5) to (7) contains NC bilinear terms (eq. 7), which is NC x NIN less than Eq. (1) to (4).

$$\sum_{i=1}^{NIN} F_{i,j} - F_{out,j} = 0 \tag{5}$$

$$\sum_{j=1}^{NC} F_{out,j} - F_{out} = 0 \tag{6}$$

$$F_{out} x_{out,j} - F_{out,j} = 0 \tag{7}$$

**Stream divider**

If a stream is divided into several (NOUT) flows, that is described by equations (7) and (8), which hold for both (total flow, fractions) and for (component flows, fractions) paradigms. Please note that in the (component flows, fractions) paradigm, it is required that the inlet stream to a flow splitter be described by both component flows and by fractions.

$$F_{in} - \sum_{k=0}^{NOUT} F_k = 0 \tag{7}$$

$$x_{in,j} - x_{k,j} = 0 \qquad j = 1, \dots, NC; \quad k = 1, \dots, NOUT \tag{8}$$

Equation (7) assumes that the outlet flows are determined from some downstream requirements. If split fractions $\alpha_k$ of the inlet stream into outlets is predefined, one needs to add eq. (9).

$$F_{in}\alpha_k - F_k = 0 \qquad k = 1, \dots, NOUT \tag{9}$$

**Component separator**

Component separator produces streams that each may have composition different from the feed stream.

*Fraction-based separator model* is described by Eq. (10) to 13). It contains NC bilinear terms (eq. 10) plus NC x NOUT bilinear terms (eq. 13),

$$F_{in}x_{in,j} - F_{in,j} = 0 \qquad j = 1, \dots, NC \tag{10}$$

$$F_{in,j}\alpha_{k,j} - F_{k,j} = 0 \qquad j = 1, \dots, NC; \quad k = 1, \dots, NOUT \tag{11}$$

$$\sum_{j=1}^{NC} F_{k,j} - F_k = 0 \qquad k = 1, \dots, NOUT \tag{12}$$

$$F_k x_{k,j} - F_{k,j} = 0 \qquad j = 1, \dots, NC; \quad k = 1, \dots, NOUT \tag{13}$$

*Flow-based component separator model* is described by Eq. (14) and (15). It does not contain any bilinear terms, which is NC x (1+NOUT) less than fractions-based separator model.

$$F_{in,j}\alpha_{k,j} - F_{k,j} = 0 \qquad j = 1, \dots, NC; \quad k = 1, \dots, NOUT \tag{14}$$

$$\sum_{j=1}^{NC} F_{kj} - F_k = 0 \qquad\qquad k = 1, \dots, NOUT \qquad (15)$$

**Chemical reactors**

Without attempting to generalize our conclusions, we present two examples of chemical reactors that are represented by linear models when using component mass flows. Both reactors, autothermal reforming (ATR) of methane to produce hydrogen and water-gas shift (WGS) reactor, are modelled well by RGibbs reactor in AspenPlus.

Production of hydrogen in ATR is represented accurately by Eq. (16), where $x$ are mass fractions of components in the reactor feed and $y_H$ is mass fraction of hydrogen at the exit, while T [K] is the reactor temperature.

$$\sum_{j=1}^{NC} a_j x_j + a_y y_H + a_T T = 0 \qquad (16)$$

Since mass flow of the feed $F_{feed}$ to the reactor is equal to the mass flow of products leaving the reactor (which is not the case if amounts are expressed in moles), one can multiple Eq. 16 by feed mass flow to obtain:

$$\sum_{j=1}^{NC} a_j F_j + a_y F_H + a_T F_{feed} T = 0 \qquad (17)$$

Engineering knowledge tells us that the reactor temperature should be kept at the highest possible temperature, i.e. reactor temperature $T_{ATR}$ is constant. Hence, the model of ATR is linear if one uses component mass flows. That is not possible if one uses moles & fractions model. Similarly, WGS reactor model is also linear if one uses component flows. Even if T is not constant there is only one bilinear term.

## 3. Energy balance models

**Energy balance models with fixed stream unit enthalpy**

Current production planning and scheduling models do not include plant energy balances. This is likely due to the fact that energy has always been a lot less expensive than the materials and CO2 emissions have not been important. Considerations of CO2 emissions and time-of-use energy pricing required that energy balance models be included in all versions of plant models.

Calculation of very accurate energy balances requires knowledge of stream temperatures, rigorous heat exchanger network models, as well as energy consumption / production by different process equipment. However, it is possible to compute accurate energy balances as long as the process streams are at their typical operating conditions. For given feed composition, most of the time streams leaving process units are at specified pressure, temperature and composition that tends to remain more or less constant. The change advocated by this work is that accurate energy balances can be calculated by using enthalpy per unit mass and heat of vaporization per unit mass at typical conditions of each stream ($H^0$) and by using, and heat of reaction per unit mass at typical conditions for reactors.

A prototypical chemical plant is comprised of a feed system, reactor, separator, and a recycle of unreacted reactants (Fig. 2). Let us assume that under given temperature $T_s^0$ and pressure $P_s^0$ of each stream $s$ we can calculate unit enthalpy $H_s^0$ [energy/mass] (e.g. from a rigorous thermophysical properties software or from a rigorous plant model). We use mass instead of moles for the following reasons:

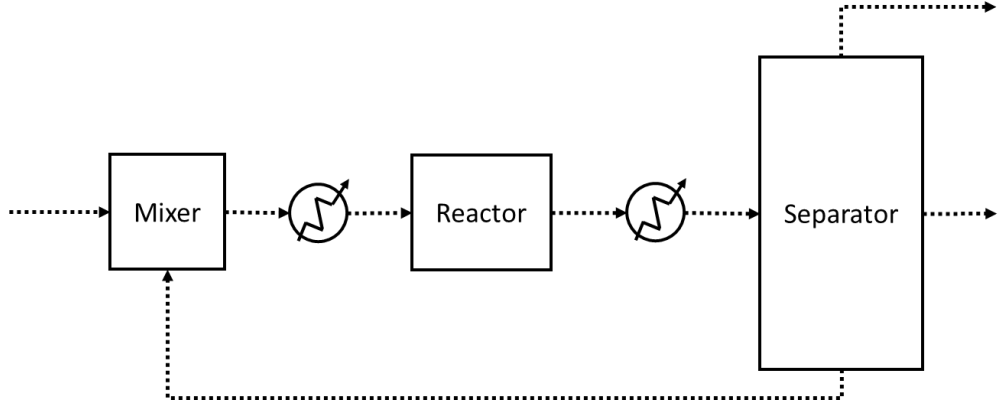

Figure 2. Prototypical process plant

Process plants measure stream flows in volumetric or mass units, not in moles. Data driven models can be developed from measured plant data and will represent the plant as it is. If one wants to develop mole-based model, that requires estimation determination of the stream composition and then conversion to moles. An unnecessary complication which also introduces additional errors.

Thermal properties (heat capacity, heat of vaporization) expressed as [energy/mass] are less sensitive to changes in composition expressed in mass units than in moles.

Energy balance for any node $n$ can then be written as:

$$\sum_{i=1}^{NIN} F_{i,n} H_i^0 + Q_{in,n} + Q_{rct,n} - \sum_{k=1}^{NOUT} F_{k,n} H_k^0 - Q_{out,n} = 0 \tag{18}$$

where $Q_{in,n}$ and $Q_{out,n}$ is energy added or removed to/from the unit, while $Q_{rct,n}$ is heat of reaction, if there is a reaction in the unit.

Such energy balance model corresponds to typical plant conditions. As long as the changes in the plant are due to changes in flows, these energy balance models are as accurate as the models built on rigorous first principles equations. Hence, using this kind of energy balance in production planning or plant scheduling enables accurate optimization w.r.t. time of use energy pricing without introducing nonlinearities.

If process unit models represent the impact of key operating variables (e.g. conversion in a reactor or different modes of operation), then mass balance models (Eq. 1 to 17) and energy balance models (Eq. 18) for each node enable optimization of plant throughput and its operating conditions without introduction of plant-level nonlinearities caused by fractions and moles stream abstraction. These types of models are required to optimize production plans and schedules of plants that consume significant amounts of energy, particularly if energy prices change frequently (e.g. electricity supplied to a blue hydrogen autothermal reforming plant). For most production planning and scheduling decision making, models represented by Eq. (1) - (18) account as accurately as needed for energy consumption and production. Such models are appropriate energy intensive processes, e.g. blue hydrogen plan equipped with CHP which produces electricity for oxygen production via air separation and for hydrogen liquefaction, and possibly exporting electricity to the grid. Grid electricity prices often change every 10 to 15 minutes, which means that the optimal operations over the next several time periods need to be calculated within seconds and then promptly implemented in the plant. The requirement that a multiperiod model needs to be optimized within seconds makes it infeasible to use moles & fractions-based software (e.g. AspenPlus).

**Energy balance models that include changes in stream enthalpy**

Optimization of operating conditions may lead to changes of temperatures of the streams leaving some process units. Such changes are limited in magnitude since the process unit (e.g. reactor) is still expected to produce the same product, albeit with some variations in product composition or some of the product properties. If the changes in the temperatures of the outlet streams are very large (e.g. 50 K or 100 K), it can be argued that the process unit is either not longer functioning properly or that it operates in a different operating mode; the latter case can then be modelled by two models, one representing each mode of operation).

Since changes in the stream temperatures are limited, unit enthalpy for each stream can be updated by:

$$H_k = H_k^0 + C_p(T_k - T_k^0) \tag{19}$$

Fig. 3 depicts the node and stream abstraction levels, depending on calculation of stream unit enthalpy.

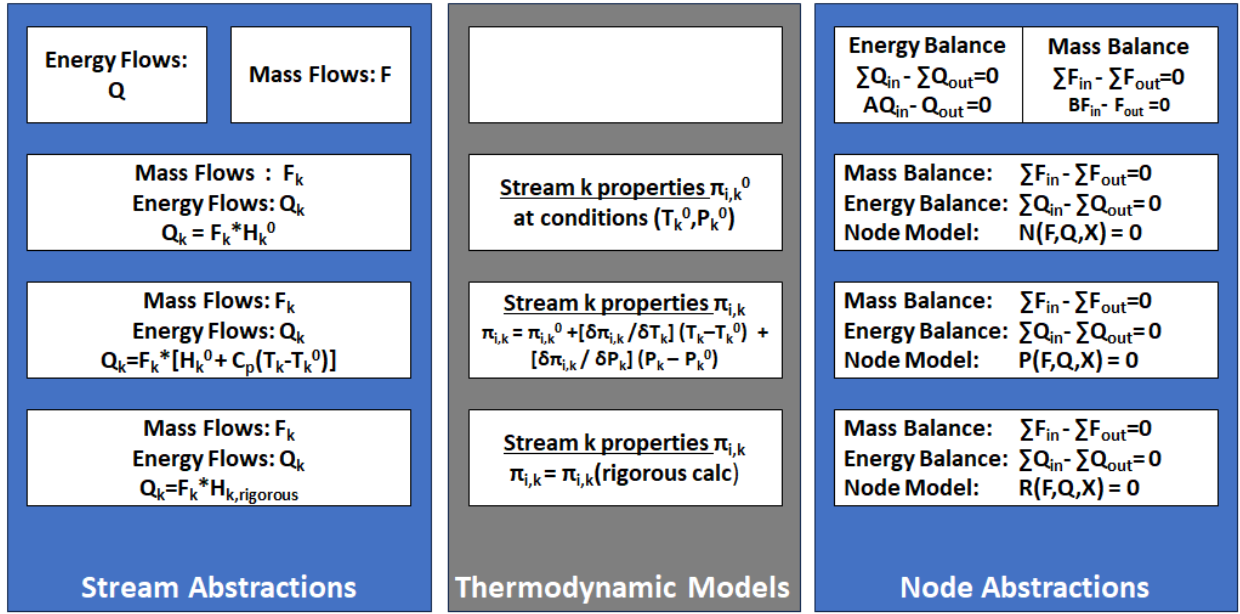

Figure 3. Stream and node mass and energy abstraction levels

**Heat exchanger networks**

Solution of mass balance and energy balance with fixed stream unit enthalpies, Eqs. (1) to (18), does not consider changes in temperatures of the streams which may occur if there are heat recovery exchangers in the plant. If stream temperature changes due to presence of heat exchangers need to be included in the model, (Ijaza 2013) presented an algorithm which solves rigorously heat exchanger networks (based on (Goyal 1975) who reformulated heat exchanger LMTD equations):

1. Determine the flows through the network via mass balance set of equations – Eq. (1) to Eq. (17) and energy balances with fixed stream unit enthalpies, Eq. (18). Note that the nonlinear terms will appear only in stream dividers and possibly in reactor models.

2. For each heat exchanger, calculate heat exchanger duty ratio $\phi$ relative to the exchanger duty $Q_{base}$ at some known conditions. This ratio depends only on the flows through the exchanger and the conditions corresponding to the base duty.

3. For each exchanger, the duty corresponding to the flows calculated by (1) - (17) is then given by Eq. (20)

$$Q_n = \phi_n \, Q_{base} \left( T^{in}_{hot,n} - T^{in}_{cold,n} \right) \tag{19}$$

4. The next step is to solve energy balance equations (20), and (21) to compute the stream temperatures from energy balances for hot and cold streams of each exchanger and energy balances for the remainder of the nodes in the plant model. Note that Eq. (20) and (21) are liner equations since exchanger duty and flows are known at this point.

$$Q_{hot,n} = F_{hot,n} H^{0,in}_{hot,n} + Cp_{hot,n} \left[ \left( T^{in}_{hot,n} - T^{0,in}_{hot,n} \right) - \left( T^{out}_{hot,n} - T^{0,out}_{hot,n} \right) \right] \tag{20}$$

$$Q_{cold,n} = F_{cold,n} H^{0,in}_{cold,n} + Cp_{cold,n} \left[ \left( T^{in}_{cold,n} - T^{0,in}_{cold,n} \right) - \left( T^{out}_{cold,n} - T^{0,out}_{cold,n} \right) \right] \tag{21}$$

5. The new value of the stream temperatures can now be used to update stream unit enthalpies (from Eq. (19) or from rigorous thermodynamic calculations) and repeat the calculations from step 1 until changes of stream temperatures are within tolerance. Ijaz et al (2013) also describe how to model exchangers with phase change.

## 4. Consistency of solutions at different abstraction levels and from one decision making process to another

If all incarnations of plant models are instantiated from the same plant topology as illustrated in Fig. 4, then moving from one abstraction level to another means that the solution from a higher abstraction level is available as a starting point for a solution of the more detailed abstraction.

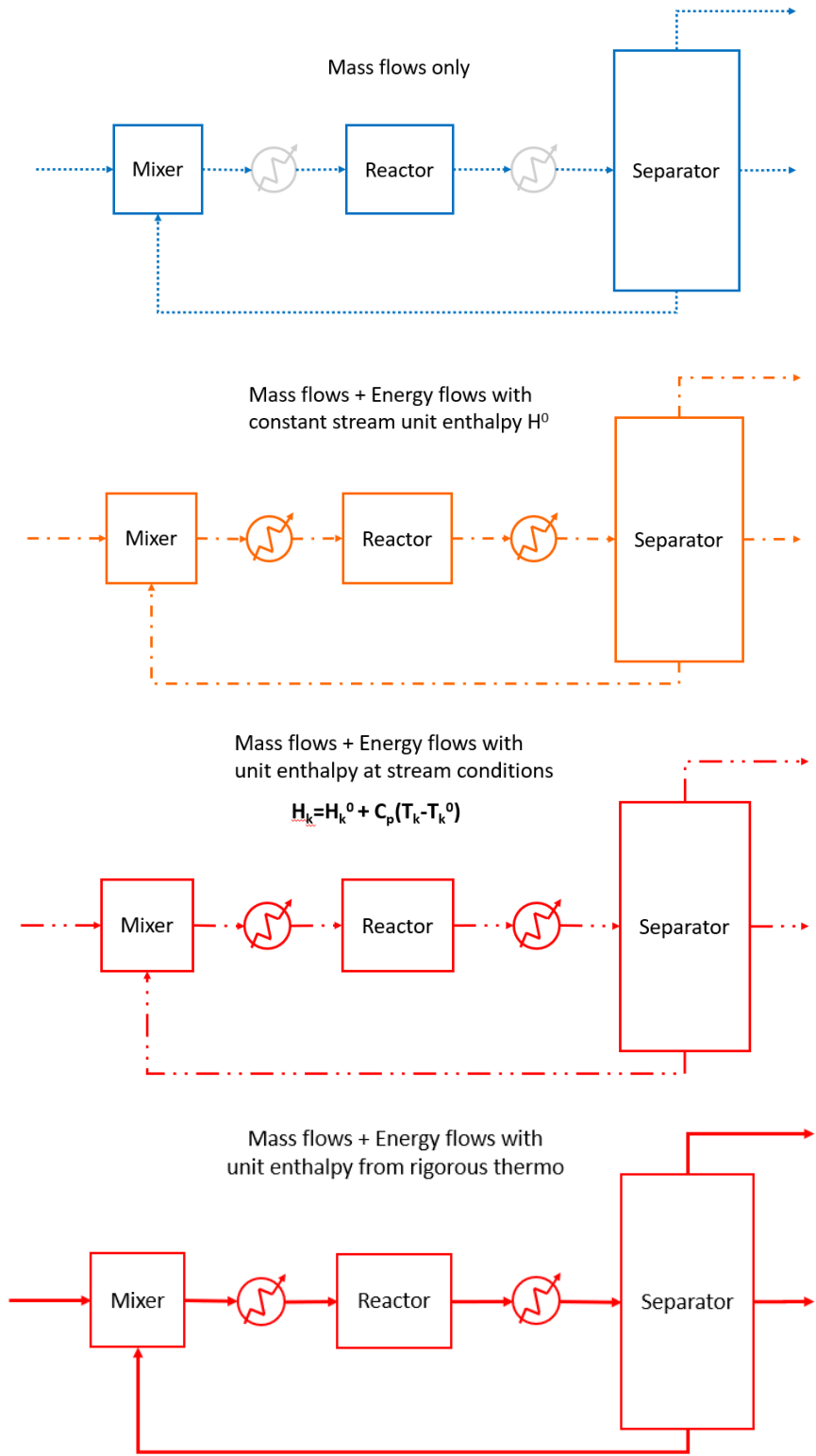

Figure 4. Consistency among abstraction levels is based on the same plant topology

Please note that all node models, e.g. heat exchangers, incarnation as "mass balance only" have no energy balances of energy transfer equations.

## 5. Solution algorithms

If all models for decision making in the plant are instantiated as different incarnations from the same plant topology, then solution from one decision-making step corresponds to the solution in another decision-making step, e.g. solution of a scheduling model, at some points along the scheduling horizon, corresponds to the planning solution along the same time horizon. Fig. 5 maps different abstraction levels to plant models for different decision-making processes.

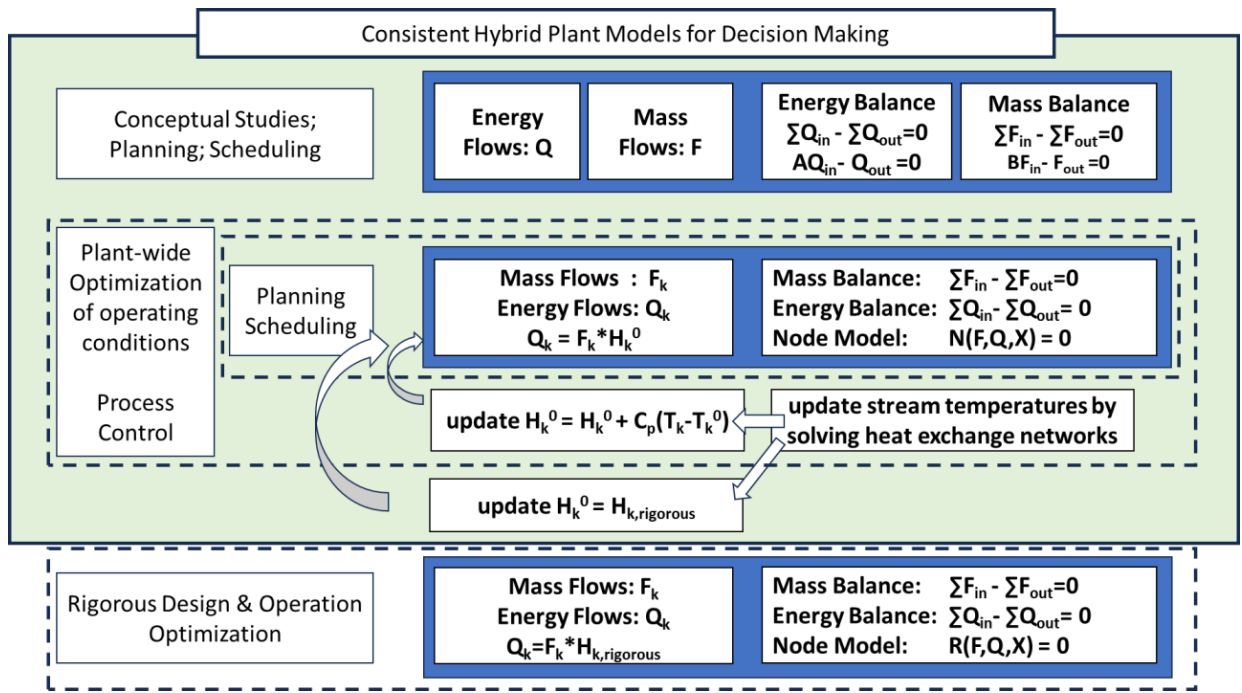

Figure 5. Mapping of plant decision making processes to abstraction levels of a plant model

Instead of solving all model equations at once, let's construct algorithms that will approach the optimum in multiple steps. The first big step is to optimize the plant model described by mass balances and energy balances described by the stream unit enthalpy (and other properties) at typical conditions of each stream. Solution of such a model will lead us to the proximity of the optimum. Such a solution is likely to be sufficiently accurate for production planning and scheduling (and it will be more accurate than the typical planning and scheduling models are today).

If process nodes change their outlet temperatures (e.g. the optimal solution may be to move a reactor to a different mode of operation, or the node is a heat exchanger without control of its outlet stream temperatures), we can update the stream unit enthalpies to the new temperatures determined by the operation of each relevant node, as depicted by the dashed frame in Fig. 6.

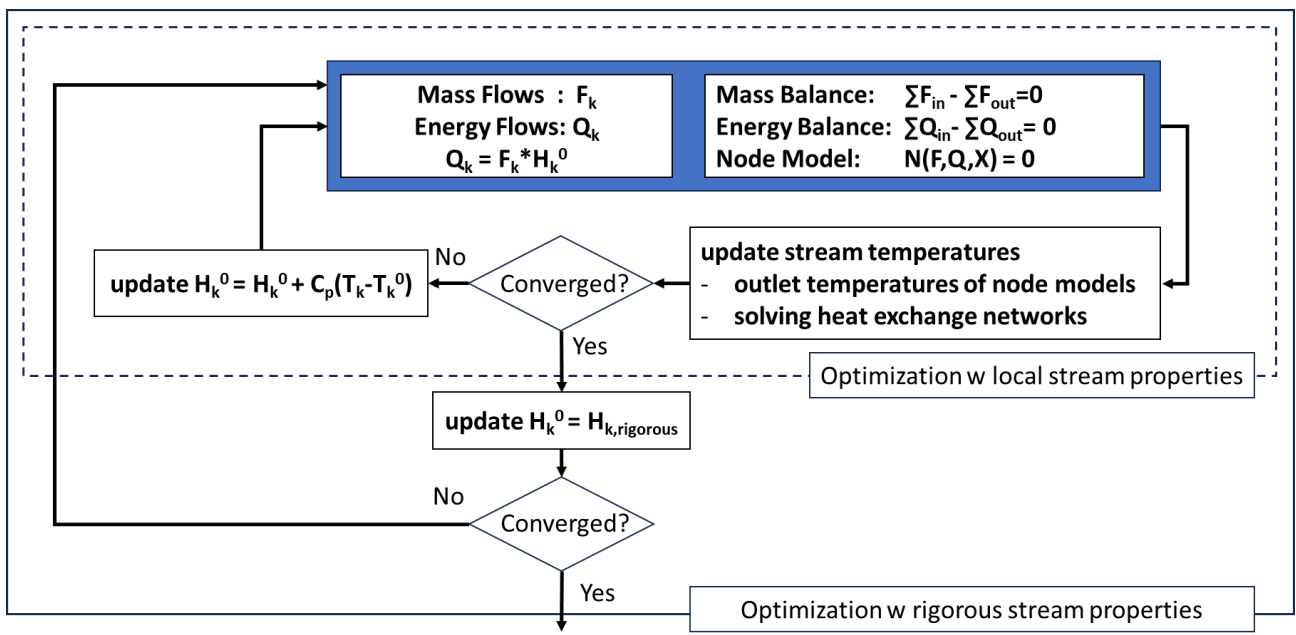

Figure 6. Convergence with local and with rigorous property calculations

If local updates of stream properties are not sufficiently accurate, we can add an outer loop where the stream properties are updated from rigorous thermophysical properties calculations. This algorithm is similar to (Boston 1978) "inside-out" flash algorithm which has excellent convergence properties. This indicates that similar convergence behavior will be observed at the flowsheet level for algorithm depicted in Fig. 6.

## 6. Plant models with different levels of abstraction at nodes, flowsheet sections, and time periods

Paradigms described until now treat entire plant models at the same level of abstraction. If we were to stay with that notion, it would simply mean that one is adding these paradigms to the three current fixed paradigms. Plant models should be instantiated at the level of abstraction that meets the needs of the business processes that will rely on the model to make decisions. Each flowsheet node, corresponding to a specific equipment, needs to be able to generate model equations which correspond to the level of abstraction of that is required for specific decision making. It is not necessary to model all parts of the plant at the same level of detail. For instance, in order to achieve the desired level of accuracy, a heat exchanger and a separator may be modelled by rigorous first principles models, while the remainder of the plant may be modelled by hybrid node models and stream enthalpies at normal operating conditions, as illustrated in Fig. 7.

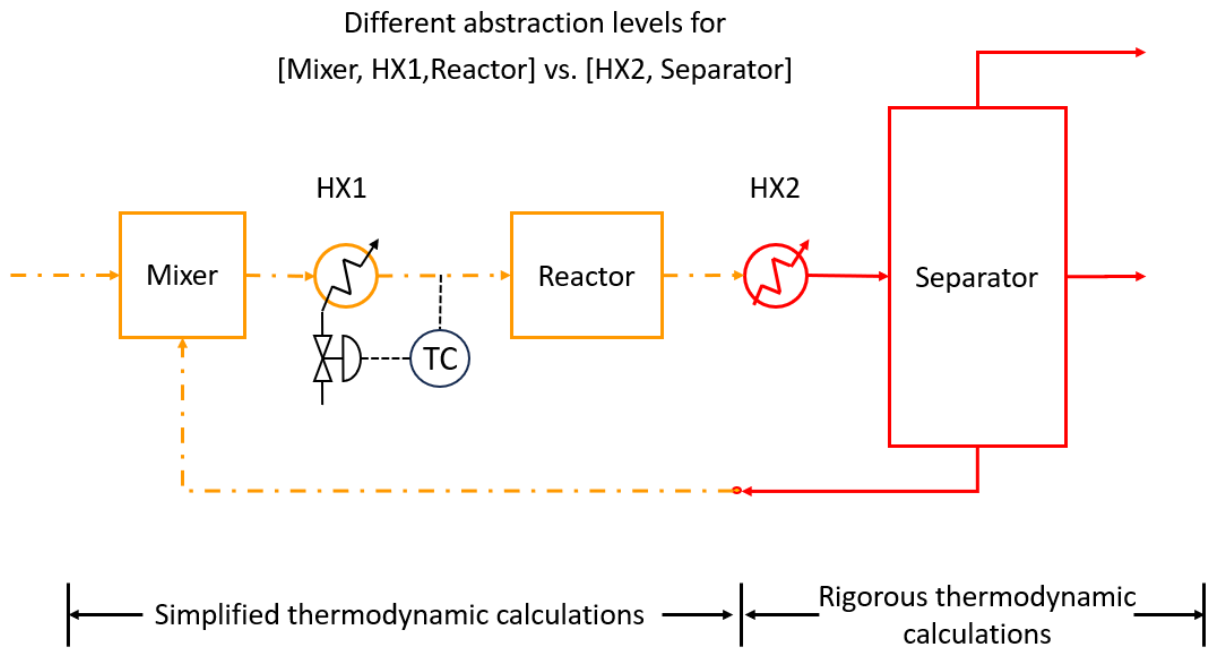

Figure 7. Each of the plant sections modelled at different level of abstraction

The ability to model different sections of a plant and in different time periods at different abstraction levels creates a possibility to approach solving some vexing business issues in a novel manner, as illustrated below.

**Integration of real-time optimization and control**

Most real-time optimization (RTO) applications consider plant operation at a given point in time, which can lead to RTO making decisions that are not optimal over some time horizon, e.g. over the next shift or the next couple of days. RTO based on rigorous first principles models is hampered by the sheer size and the nonlinearities of these models.

Let's consider a hydrogen plant which produces hydrogen from natural gas via autothermal reforming (ATR) and water gas shift (WGS) reactors. Hydrogen product is sent to a pipeline and also shipped via trucks to local consumption. A simplified representation of the plant is shown in Fig 8.

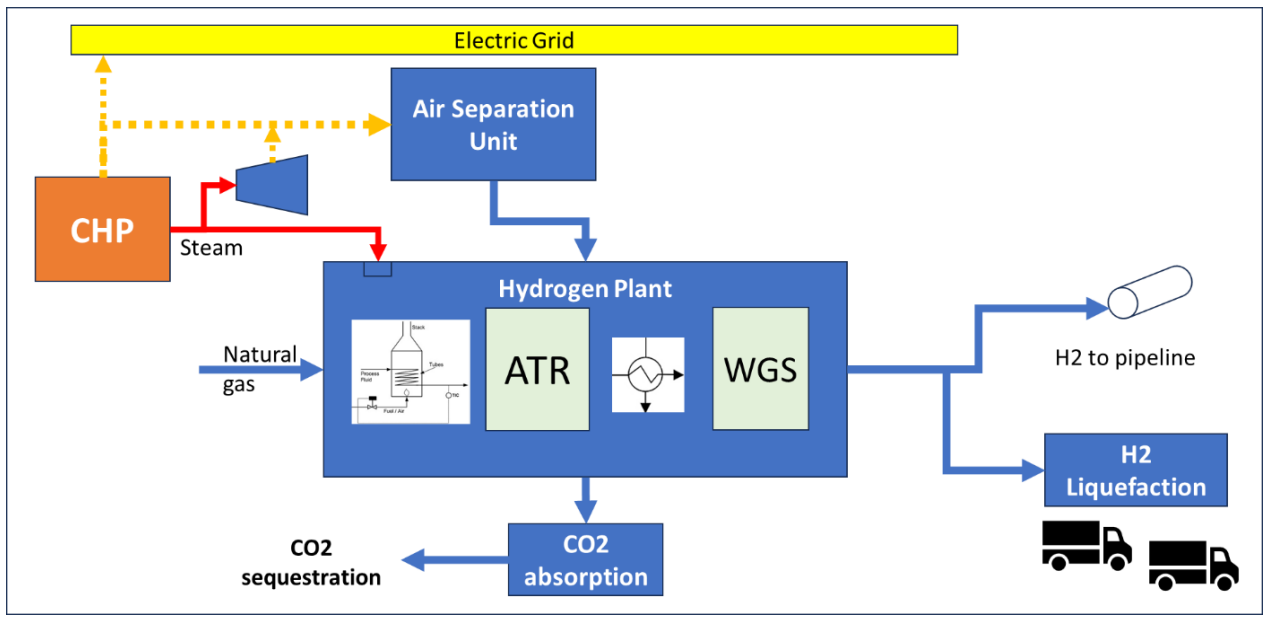

Figure 8. Production of hydrogen from natural gas

Since electricity prices change frequently (e.g. every 15 minutes), the real-time optimization of the plant operation needs to consider operation over a time horizon that may span the time required for the material to flow through the plant (e.g. 1 hr), instead of just a point in time, which leads to a model of with at least 4 periods. In order to avoid a mismatch between RTO and the control model, we may choose to implement the plant models in for the first 4 periods as hybrid models comprised of MPC model augmented by mass and energy balances (see Fig. 9), while subsequent periods may use steady-state hybrid models. If the plant models are described by component mass flows, energy balances, and reactor models as described above, the six time-period model will have only a handful of bilinear terms, which will provide robust, rapid convergence. If stream temperatures change as a result of optimization, then a composite algorithm described in Section 5 can be used to optimize the entire multi-period model.

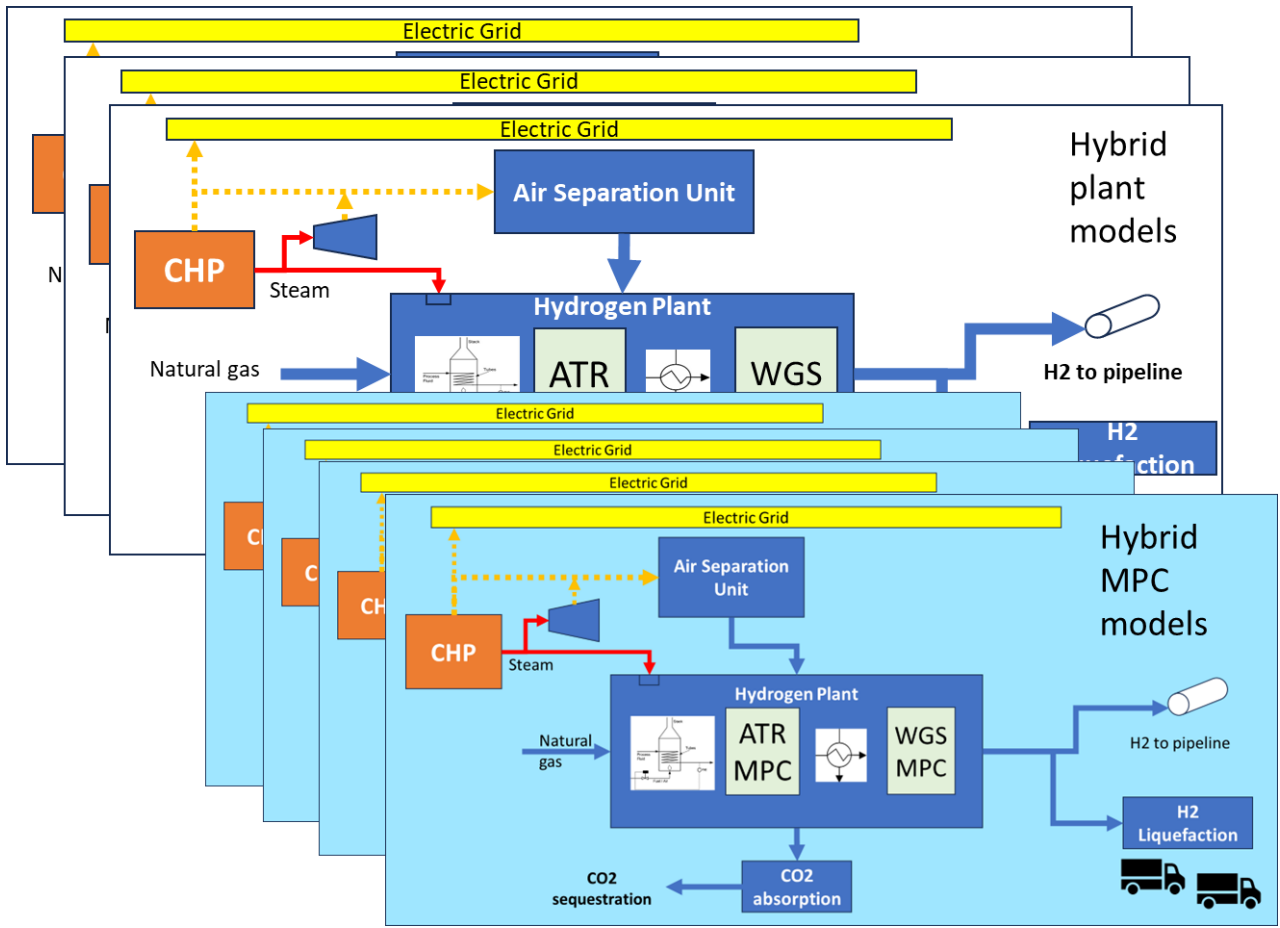

Figure 9. Integrated real time optimization and process control

If there is a plant modelling software that makes it possible to instantiate different levels of details of the plant models in different time periods, the proposed approach to solving integrated RTO and control will be very simple to implement. Moreover, the proposed approach is a scalable alternative to dynamic RTO, and it is suitable for dynamic real time optimization of entire plants.

**Integration of scheduling and control**

Scheduling of process plant operation has been approached via discrete-time models and continuous-time models. While discrete-time models are not as precise as continuous-time models, they provide faster computations of schedules for process plants. In industrial practice, scheduling has been more widely applied in batch plants, while in multi-product continuous plants (e.g. petroleum refineries) scheduling is still done by verifying heuristic schedules via simulation.

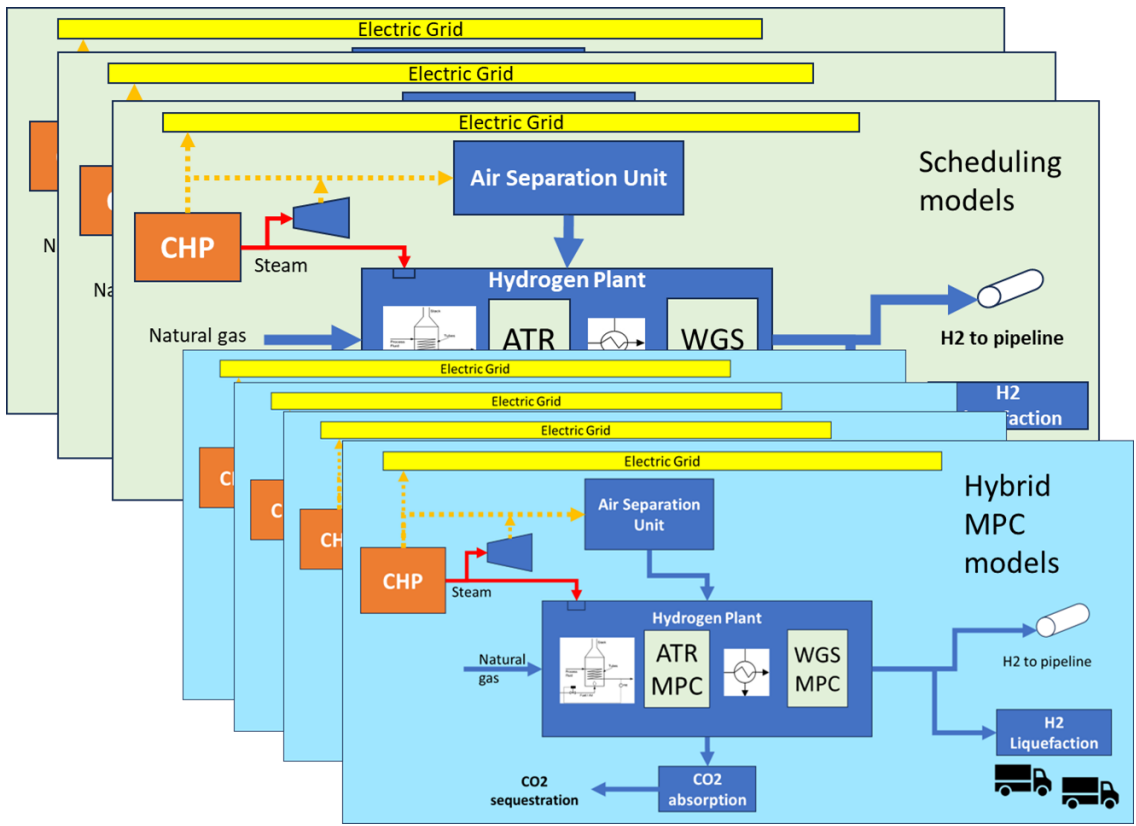

Figure 10. Integration of scheduling and control

Since scheduling models include binary variables, it is required to represent the plant by a linear model to achieve acceptable computation times. Plant models based on component mass flows and unit enthalpy fixed at local stream conditions, together with linearized representation of reactors and separation units, are more accurate than the current scheduling models, and if they are implemented in the proposed paradigm, then the scheduling models inherit solutions from the longer-term production planning models.

It is proposed that the scheduling and control can be integrated via a multi-period model, where in the initial periods the plant is modelled by hybrid MPC models (models comprised of mass an energy balances and of MPC process models), while the later periods are scheduling models as described above (Fig. 10).

### 7.　　Why we need a new generation of software

Current software is rooted in three different paradigms that originated decades ago. Each paradigm uses different plant representation (different plant topology) and different levels of abstraction without inheritance from one abstraction to another. That causes inconsistencies between decision making steps, extensive (mostly unsuccessful) efforts to eliminate these inconsistencies, and limits quality of the solutions that are obtained.

Common to all three paradigms is adherence to a single abstraction level and to single pass solution algorithms, i.e. generate all model equations at once and solve them. Consequences of employing single abstraction level, based on different plant topologies, to the entire plant model are:

- First principles plant models typically have tens of thousands of nonlinear equations which limits their use to a single time period and optimization times that are often not acceptable real-time business decision making.
- Planning and scheduling models can not:
    - represent different sections of a plant at different level of abstraction
    - employ different levels of abstraction in different time periods

- why should the models representing the next week be at the same level of accuracy as the models representing a time period that is 10 months away?
- MPC models do not include mass and energy balances and cannot be integrated into tim horizons covering several days.
- Solutions between planning, scheduling, RTO, and control are inconsistent.

With current generation of software, it is inherently not possible to construct composite algorithms, for instance:

1. Solve multiperiod mass balance and fixed stream unit enthalpy model (i.e. planning model)
2. Solve heat exchanger network model for the flows from 1. above.
3. Update stream enthalpies at the new stream temperatures.
4. Solve multiperiod mass and energy balances with updated stream unit enthalpies (i.e., plant wide optimization)
5. Etc.

Unfortunately, none of the existing plant modelling software (AspenPlus, PRO/II, ROMEO, Aspen HYSYS, Aspen PIMS, Aspen Petroleum Scheduler, Aspen Unified, etc.) have capabilities required to build plant models as described here.

There has been some recognition that different abstraction levels are required to solve efficiently large-scale models. For instance, the latest generation of production planning and scheduling (Aspen Unified) allows different time-period lengths but does not permit different levels of stream or node abstractions in different plant sections or in different time periods. While this is a steps in the right direction, it should be noted that diverse time-period granularity is not sufficient, it is a minor part of the required multitude of plant model abstractions.

## 8. Conclusions

Plant model incarnations from the same process topology is the key to ensuring consistency among various plant models. The proposed paradigm inherently enables solutions from one level of abstraction to be consistent with solutions at another level of abstraction, i.e. from one business process to another.

Switching from (mole & fractions, rigorous properties) to (mass component flows, properties at the standard stream conditions) enables comparable level of accuracy while eliminating nonlinearities associated with stream state calculations. This makes it possible to build planning and scheduling models that are linear (with respect to stream representation) and yet as accurate as the rigorous models at the typical plant operating conditions.

Optimization of plant operating conditions based on hybrid models can be improved by local approximations of stream physical properties. Mass-based physical properties are less sensitive to changes in composition than mole-based physical properties. In addition, using mass units makes it easier to employ data-driven process models, since measurements in the plants are either in volumetric or mass units but not in moles.

Component mass flows (instead of mole fractions) open a possibility that hybrid reactor models can be less nonlinear than mole-fraction based reactor models.

Different node model abstraction levels, in different sections of a plant model and in different periods, open gates to new approaches to optimal planning, scheduling, RTO, and control; to their integration and ensuring consistency of solutions between them.

Finally, the concepts presented in this paper raise a question whether the current moles & fractions paradigm is the best choice for rigorous, first-principles based modelling of process plants. If we imagine that instead of hybrid node models one is to employ the first-principles node models while keeping the stream representation as component mass flows, would we end up with much better convergence properties for the entire process flowsheet?